\documentclass[aps,prl,twocolumn,amsmath,amssymb,floatfix,superscriptaddress]{revtex4}
\usepackage[T1]{fontenc}
\usepackage[latin1]{inputenc}
\usepackage{amsmath}
\usepackage{graphicx}
\usepackage{epsfig}
\usepackage{bm}
\newcommand{\nn}{\nonumber\\}

\newcommand{\bea}{\begin{eqnarray}}
\newcommand{\ea}{\end{eqnarray}}
\newcommand{\beq}{\begin{equation}}
\newcommand{\eq}{\end{equation}}
\newcommand{\bc}{\begin{center}}
\newcommand{\ec}{\end{center}}
\begin{document}

%\psdraft
%\title{Coherent particle oscillations between two Bose-Einstein condensates 
%mediated by a single localized impurity atom}
\title{Coherent single atom shuttle between two Bose-Einstein condensates}

\author{Uwe R. Fischer}

\affiliation{Eberhard-Karls-Universit\"at 
T\"ubingen, Institut f\"ur Theoretische Physik,  
Auf der Morgenstelle 14, D-72076 T\"ubingen, Germany}

\author{Christian Iniotakis}

\affiliation{ETH Z\"urich, Institut f\"ur Theoretische Physik, 
CH-8093 Z\"urich, Switzerland}

\author{Anna Posazhennikova}
%$^{\dagger\,{\scriptstyle 1}}$}

\affiliation{Physikalisches Institut, Universit\"at Bonn, D-53115 Bonn, 
Germany}  
\affiliation{Institut f\"ur Theoretische Festk\"orperphysik and DFG-Center
for Functional Nanostructures,
Universit\"at Karlsruhe,
D-76128 Karlsruhe, Germany}

%\preprint{version of \today}

\date{\today}

\begin{abstract}

We study an atomic quantum dot 
representing a single hyperfine ``impurity'' atom which is 
coherently coupled to two well-separated Bose-Einstein condensates, 
in the limit when the coupling between the dot and the condensates dominates
the inter-condensate tunneling coupling. 
It is demonstrated that the quantum dot by itself can induce 
large-amplitude Josephson-like 
oscillations of the particle imbalance 
between the condensates, which display a two-frequency behavior.
For noninteracting condensates, we provide an approximate 
solution to the coupled nonlinear equations of motion which allows us to 
obtain these two frequencies analytically.

\end{abstract}
\maketitle

%The fundamental property of the ordered, broken-symmetry
%states of superconductors, 
%superfluid $^3$He or $^4$He, and Bose-Einstein
%condensates of dilute atomic gases (BECs) is their phase coherence.
When two phase-coherent quantum systems are brought close
together, but are still separated by a 
tunneling barrier, particle current oscillations are induced. 
The exploration of the Josephson effect, %in the form originally 
predicted in 1962 \cite{Josephson}, 
was for many years limited to superconducting
materials, until in 1997 quantum oscillations through an array of weak 
links in superfluid $^3$He-B  were observed \cite{Pereverzev}.
%, realizing Josephson tunneling in an electrically neutral system.

Dilute atomic gases, which upon lowering the temperature to the 
sub-$\mu$K-range %Bose-Einstein-condense and therefore 
acquire phase coherence, allow to 
%provide a new and promising playground for 
study macroscopic coherence effects in a highly controlled manner, with 
essentially single-atom accuracy. 
% and under exotic circumstances. 
%As a distinguishing feature, the tunneling particles 
%in the Josephson effect interact with themselves. 
Calculations based on the two-mode description 
\cite{MilburnJosephson}
of two weakly coupled condensates \cite{Smerzi,Zapata} 
have exemplified the 
rich dynamics of particle imbalance oscillations inherent to 
Josephson junctions between BECs. In particular, upon increasing the particle interaction, two weakly-coupled condensates enter the novel collective  
state of macroscopic self-trapping, which is due to the 
self-interaction of the tunneling particles \cite{Smerzi,Raghavan}, and 
has been confirmed experimentally \cite{Albiez,Levy}.

In the following, we explore the indirect particle exchange between two 
condensates, mediated by a single ``impurity'' atom coherently 
coupled to the two condensates, which is located in a tight trapping
potential at the position of the barrier 
between the condensates. 
The conventional 
tunneling channel we assume to be strongly suppressed by raising the barrier 
between the two condensates.  
%The impurity consists of a single atom
%of a hyperfine species different from that of the condensates.  
The dynamics of the type of impurity we consider -- an atomic quantum dot 
(AQD) -- when coupled by laser transitions to a single infinite
superfluid reservoir, was studied in \cite{Recati}. When the AQD is coupled to
two well-separated condensates, it was demonstrated in \cite{Bausmerth} that 
when the dot-condensate coupling is smaller than or comparable to 
the intercondensate tunneling 
coupling, the effect of AQD on the particle oscillations 
between the condensates is negligibly small.
%, in accordance with what was  obtained for superconducting junctions
% \cite{Shnirman}.
The specific question we address in the present work is if the 
single impurity AQD can act as a {\em coherent} ``shuttle'' 
between the essentially isolated BEC reservoirs when particle transfer 
between wells is possible only via the coupling of the dot to the 
condensates: We term this the ``strong coupling'' limit. 
We find unexpected behaviour in this strong
coupling limit; namely, apart from expected 
small and rapid oscillations of the particle
imbalance due to single particles going to and fro between 
dot and condensates, large-amplitude Josephson-like 
oscillations between the condensates, mediated by the dot,  
occur at a smaller frequency.

%, transferring atoms from 
%left to right and vice versa such that effective Josephson oscillations 
%between the two wells are established.  
%We demonstrate that even in the limit of vanishing 
%conventional tunneling, macroscopic 
%coherence between the two wells is 
%established by the single coherently 
%coupled atom in the AQD. 

Our model system consists of two condensates with a large 
(single-particle) potential
barrier erected between them, 
and a single impurity atom,  
coupled to the condensates by a two-photon Raman transition. 
%At the end of the paper, we discuss a possible concrete
%experimental implementation of this theoretical model first proposed
%in \cite{Bausmerth}. 
Assuming a symmetric double-well 
trap, the system is described by the following 
Hamiltonian, %in single-occupation limit with all overlap integrals neglected 
\bea 
\hat H &=& %E^0_i |\varPsi_i(t)|^2 
U \left[ |\varPsi_1(t)|^4 + |\varPsi_2(t)|^4\right]
-\kappa \left[\varPsi_1^*(t)\varPsi_2(t)+{\rm h.c.} \right]  
\nn 
&  & 
+ T\sum_{i=1,2}\left\{  \varPsi_i (t)  \hat \sigma_+  +{\rm h.c.}  \right\}
-\hbar\delta \frac{1 + \hat \sigma_z}2.  \label{H} 
\ea 
The above Hamiltonian is valid within 
the two-mode approximation \cite{MilburnJosephson} 
for the total condensate wave 
function, $\varPsi({\bm r},t)
=\varPsi_1(t)\phi_1({\bm r})+\varPsi_2(t)\phi_2({\bm r})$, 
where the single-particle wave functions $\phi_{1,2}({\bm r})$ 
(normalized to unity) describe the
particles localized in their respective wells, 
and $\varPsi_{1,2}(t)$ are time-dependent amplitudes representing
the tunneling process. 
The pseudospin (equivalent to the two-level system represented by the dot) 
is defined by the Pauli matrix vector $\hat{\bm \sigma}(t) 
=(\hat \sigma_x,\hat \sigma_y, \hat \sigma_z )$,  
%$ {\bm s} (t)=\langle\varPsi_d(t)| \hat{\bm \sigma} |\varPsi_d(t)
%\rangle=\langle\varPsi_d| \hat{\bm \sigma}(t) |\varPsi_d \rangle.$, 
and $\hat\sigma_+ = \frac12 (\hat \sigma_x+i\hat \sigma_y) 
= \hat \sigma_-^\dagger$ is a spin ladder operator.  
The ``on-site'' interaction between the particles is  
given by $U_{i}=g\int d{\bm r}\,|\phi_{i}({\bm r})|^4 $,   
$\kappa=-\int d{\bm  r} \,[\frac{\hbar^2}{2m}(\nabla\phi_1({\bm r})\nabla\phi_2({\bm r}))
+\phi_1({\bm r}) V_{\rm \scriptscriptstyle BEC}({\bm r})
\phi_2({\bm r})]$  denotes the positive tunneling 
coupling \cite{Smerzi}, and $\delta$ is 
the detuning from the two-photon Raman 
transition coupling a single atom into the dot.
The corresponding coupling parameter (transfer matrix element) is 
$T=\hbar \Omega_R\int d{\bm r}\, \phi_1({\bm  r})\phi_d({\bm r})
=\hbar \Omega_R\int d{\bm r}\, \phi_2({\bm r})\phi_d({\bm r})$, where
the spatial wave function of the dot $\phi_d({\bm r})$ is normalized to unity, 
and $\Omega_R$ is the Rabi frequency of the two-photon Raman transition.
Note that while the overlap integrals of dot-condensate
and condensate-condensate wave functions are both small and 
comparable in order of magnitude, 
the strong coupling limit of $T/\kappa\rightarrow\infty$ 
can be achieved by sufficiently increasing the Rabi frequency $\Omega_R$, 
i.e., well above the BEC single-particle energies in the overlap region, 
which enter the tunneling coupling $\kappa$.

%### Dieses Zeichen >
From the Hamiltonian (\ref{H}), 
we derive the coupled equations of motion for the condensate
variables $\varPsi_{1,2}$  and the pseudospin vector 
${\bm s}=\langle\varPsi_d(t)| \hat{\bm \sigma} |\varPsi_d(t)\rangle$,  
respectively $s_{\pm},s_z$, of the AQD; here, $|\varPsi_d(t)\rangle$ is the 
temporal dot wave function. The condensate equations read ($\hbar \equiv 1$)
\begin{eqnarray}
i\partial_t\varPsi_1 = U |\varPsi_1|^2 \varPsi_1  
-\kappa\varPsi_2 + T s_- , \nonumber \\ 
i\partial_t\varPsi_2 = U |\varPsi_2|^2 \varPsi_2  
-\kappa\varPsi_1 + T s_- , \label{eq_cond} 
\end{eqnarray}
while the equations of motion for the pseudospin are
\begin{eqnarray} 
i \partial_t s_- &=&
-\delta s_- - T(\varPsi_1+\varPsi_2)s_z, \nonumber 
\\
i \partial_t s_z &=&  -2T( \varPsi_1^*+ \varPsi_2^*)s_-
+2T(\varPsi_1+\varPsi_2)s_+.\label{eq_dot}
\end{eqnarray} 
We now scale time with $T^{-1}$, %(which has units of energy), 
and introduce the following set of dimensionless control parameters
\bea
\alpha =\frac{UN_0}T , \quad \beta =\frac\delta T,\quad \Gamma= \frac\kappa T,
\ea
where $N_0=N_1(0)+N_2(0)=|\varPsi_1(0)|^2 + |\varPsi_2(0)|^2$ is the 
sum of the initial number of particles in the left and right 
wells, respectively; we also employ the scaling 
$\varPsi_i \rightarrow \varPsi_i/\sqrt{N_0}$. 
Decomposing $\varPsi_i$ and $s_\pm$ 
into their real and imaginary parts,
we then obtain seven 
equations for the coupled motion of the two 
condensates and the AQD pseudospin, which we solve numerically. 

\begin{figure}[!hbt]
\begin{center}
\includegraphics[width=0.32\textwidth]{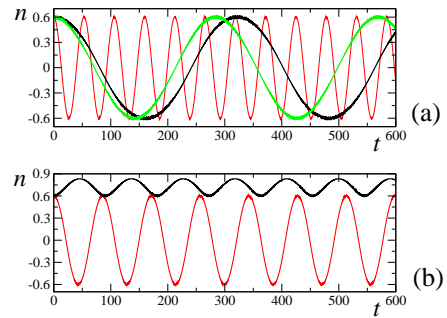}
\end{center}
\caption{\label{large} (Color online)
Large amplitude Josephson-like oscillations around zero particle imbalance, 
$n=0$, 
for $N_0=1000$, $\beta=10$, initial $n(0)=0.6$ and initial phase difference $\phi(0)=\pi$ for $\alpha=0.01$ in 
(a) and $\alpha=0.1$ in (b) [where crossover to a self-trapped state has 
occured for $\Gamma =0$]. Black solid curve: $\Gamma=0$, red (thin grey) curve:
$\Gamma=0.05$. %(the critical $\alpha_c \sim 0.028$ for $\Gamma=0$). 
In (a), we display in addition the noninteracting case ($\alpha=0$)  
with $\Gamma=0$ in green (thick grey). We assume throughout our calculations 
that there is initially one particle in the dot, $s_z(0)=1$ and that the 
initial particle imbalance $n(0)=0.6$; time is in units of $T^{-1}$.}
\end{figure}

\begin{figure}[!hbt]
\begin{center}
\includegraphics[width=0.3\textwidth]{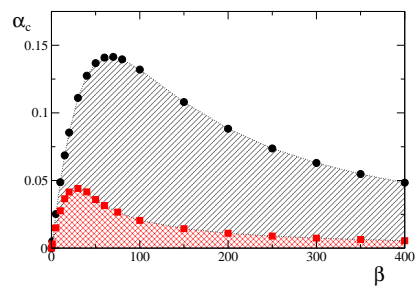}
\end{center}
\caption{\label{phase} (Color online) The critical value of the scaled interaction 
$\alpha=\alpha_c$ for self-trapping of the effective Josephson oscillations, 
as a function of the scaled detuning $\beta$, for a $\phi(0)=0$--junction 
(black dots) and  $\phi(0)=\pi$--junction (red squares), 
with $N_0=1000,\Gamma=0, 
n(0)=0.6$. Within the shaded areas, the system shows Josephson-like 
oscillations, while inside the white area, it is self-trapped.}
\end{figure}

We first concentrate on the strong coupling case $\Gamma \rightarrow 0$.  
%by raising the barrier to very high values
%(amounting to $\Gamma$ being very small, i.e., of the order of
%$10^{-4}$ or smaller).  
Our major result is that large amplitude 
Josephson-like oscillations of the particle imbalance
$n(t)=(N_1(t)-N_2(t))/N_0$, with amplitude $n(0)$, can be induced by the
quantum dot, which can coherently transfer
atoms one by one from left to right and vice versa even when
conventional tunneling is completely switched off (Fig.\ref{large}\,(a), 
black curve). With increasing $\alpha$, macroscopic self-trapping, defined by 
an average $\langle n(t) \rangle \neq 0$, occurs (Fig. \ref{large}\,(b), 
black curve). Thus, there is a
critical value $\alpha_c$, depending on $\beta$, such that for
$\alpha>\alpha_c$ particle imbalance oscillations are
self-trapped and for $\alpha<\alpha_c$, Josephson-like
oscillations of the particle imbalance with $\langle n(t) \rangle =0$ occur. 
We have studied in detail the dependence of $\alpha_c$ on $\beta$ 
and present the results for 
$0$- and $\pi-$junctions in the phase diagram Fig.\,\ref{phase}.
We observe that, for small $\beta$, $\alpha_c$ increases linearly in $\beta$.
For large $\beta$, % (or small $T$, at fixed detuning $\delta$),
$\alpha_c\sim 1/\beta$, implying 
%that the effective coupling between the condensates, induced by the dot,
that the critical interaction $U_c \propto T^2/\delta$ in this limit. 

We stress that the self-trapping crossover obtained here  
is very different from the well-known one \cite{Smerzi}, 
as it occurs also if ordinary tunneling is  blocked.
In the conventional self-trapping 
scenario, the latter fact would imply
(trivially) that the system is always self-trapped, with zero oscillation 
amplitude. Here, by contrast, the coupling to the AQD can induce 
Josephson-like oscillations for sufficiently small $\alpha$, cf. 
Fig.\,\ref{large}.  To further emphasize the difference to the 
conventional scenario, varying the particle number, we find that for large
$\beta$ the critical $\alpha_c$ becomes essentially independent of $N_0$, 
while for small $\beta$, $\alpha_c$ {\em decreases} approximately 
linearly in $N_0$ with increasing $N_0$ [keeping $n(0)$ fixed], which is
opposite to what one would expect for macroscopic self-trapping 
driven by the total interaction energy. 
%This behavior further stresses the essential difference
%between the present self-trapping mechanism and the conventional one 

%\vspace*{-2.4em}
%%%%%%%%%%%%%%%%%%%%%%%%%%%%%%%%%%%%%%%%%%%%%%%%%%%%%%%%%%%%%%%%%%%%%%%%%%%%%
\bc
%\vspace*{-1em}
\ec
%%%%%%%%%%%%%%%%%%%%%%%%%%%%%%%%%%%%%%%%%%%%%%%%%%%%%%%%%%%%%%%%%%%%%%%%%%%%%
%%%%%%%%%%%%%%%%%%%%%%%%%%%%%%%%%%%%%%%%%%%%%%%%%%%%%%%%%%%%%%%%%%%%%%%%%%%%%
\bc
\ec
%%%%%%%%%%%%%%%%%%%%%%%%%%%%%%%%%%%%%%%%%%%%%%%%%%%%%%%%%%%%%%%%%%%%%%%%%%%%%

We observe that the dimensionless control parameter for the occurrence
%and magnitude 
of the Josephson-like oscillations is the ratio of the 
dot energy %(i.e. the detuning from the two-photon resonance) 
over the self-interaction energy of the condensates:
%\beq
${\beta}/{\alpha}={\delta}/{U N_0}.$ 
%. \label{gammadef} 
%\eq 
%and the Rabi frequency $\hbar \Omega_R $ is of order of GHz, 
%the large oscillations induced by the coupling to a single impurity, 
Given that the typical mean-field energy 
$U N_0 \sim$10 nK for particle numbers  
$N_0 \sim 1000$, for $\beta/\alpha =10^1\cdots 10^4$,  
the necessary detuning is of order $\delta=\frac\beta\alpha N_0 $\,[Hz], 
which is experimentally feasible.
% for the typical range we discuss, $\beta/\alpha =10^1\cdots 10^4$. 

\vspace*{-1em}
%%%%%%%%%%%%%%%%%%%%%%%%%%%%%%%%%%%%%%%%%%%%%%%%%%%%%%%%%%%%%%%%%%%%%
\bc 
\begin{figure}[!t]
\begin{center}
%\hspace*{1em}
\includegraphics[width=0.44\textwidth]{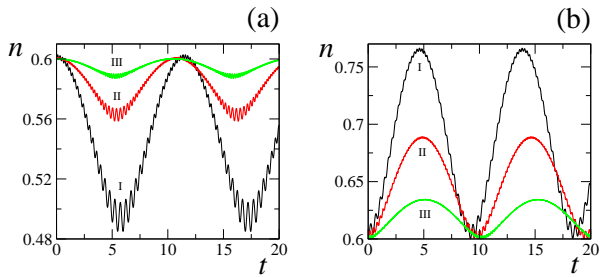}\vspace*{-0.5em} 
\end{center}
\caption{\label{small} (Color online) 
Particle number dependence of induced oscillations for 
vanishing conventional tunneling rate,  $\Gamma=0$, for 
$N_0=100$ (black, I), $N_0=300$ (red, II), and  
$N_0=1000$ (green, III). 
Parameters are $\beta=10$, $\alpha=1$, 
The initial phase difference is $\phi(0)=0$ 
in (a) and $\phi(0)=\pi$ in (b); time is again in units of $T^{-1}$.}
\end{figure} 
\ec
%%%%%%%%%%%%%%%%%%%%%%%%%%%%%%%%%%%%%%%%%%%%%%%%%%%%%%%%%%%%%%%%%%%%%

If we further increase $\alpha$ (decrease $\beta/\alpha$), 
the amplitude of the self-trapped 
oscillations becomes very small; Fig. \ref{small} (green curves). 
However, the magnitude of oscillations again increases when 
the number of particles $N_0$ is decreased. In this regime of smaller
particle numbers, one can clearly distinguish a two-frequency behavior 
of the dot-induced oscillations (which is not discernible in 
Fig.\,\ref{large}). In order to understand the origin of these two 
frequencies, we provide below an analytical derivation of the oscillations 
in the limit of noninteracting condensates.

We have also analyzed the case of a finite, but still small $\Gamma$
(Fig.\,\ref{large}, red curves). The small rapid oscillations  
on top of the envelope of the oscillations rapidly vanish if one increases 
$\Gamma$ from zero, and thus occur in the strong coupling limit only. 
Furthermore, we observe from Fig.\,\ref{large} that the change in the 
effective tunneling oscillation frequency, 
increasing the tunneling coupling $\Gamma$,
% from zero to a finite value, 
strongly depends on the value of $\alpha$.  
%(i.e., for fixed $\beta$, on the value of $\gamma$ in Eq.\,\ref{gammadef}). 
If we are in a state of ordinary oscillations around
zero population imbalance, Fig.\,\ref{large}(a), the 
oscillation frequency changes strongly; conversely, in the self-trapped 
state, we have nearly no change in oscillation frequency. 
Furthermore, from Fig.\,\ref{large}(b), we conclude that by 
a slight increase of $\Gamma$, we may switch the system from 
a self-trapped state to one with Josephson-like oscillations.

For noninteracting condensates \cite{Roati}, $\alpha=U=0$, 
an analytical approximation to the coupled equations of motion is possible,
from which we are able to reproduce the numerically established dot-induced 
oscillations in that limit. 
%### Dieser ganze Abschnitt
Defining the new variables $\psi =\varPsi_1+\varPsi_2$ and 
$\bar\psi = \varPsi_1-\varPsi_2 $, we find immediately from 
Eqs.\,\eqref{eq_cond} and \eqref{eq_dot}, %with $\alpha=0$ 
that $\bar \psi$ decouples according to 
$i\partial_t \bar\psi= \kappa\bar\psi $. Hence,
$\bar{\psi}(t)=\bar{\psi} (0) e^{-i\kappa t}$,
and we are left with the  equations 
\begin{eqnarray}
i\partial_{t}\psi & = & -\kappa\psi+2Ts_{-}, \qquad 
i\partial_{t}s_{-} = -\delta s_{-}-T\psi s_{z},\nn
\partial_{t}s_{z} & = & -4T{\rm Im}\,[\psi^{*}s_{-}]. %+2T\varPsi s_{-}^{*}.
\end{eqnarray}
These yield directly  $\partial_t(|\psi|^2 +s_z)=0$, so that
\begin{equation}
|\psi|^2 +s_z=C=|\psi(0)|^2 +s_z (0)
\end{equation} is a conserved quantity. 
Next, we write $\psi =  |\psi| e^{i\varphi}$ and $s_- =  |s_-|e^{i\lambda}$
and utilize that the variables $|\psi|,|s_-|,s_z$ are directly connected to each 
other via the conservation of both $|\psi|^2 +s_z$ and 
$4|s_-|^2 + s_z^2$. We get 
\begin{eqnarray}
%\partial_{t}(\phi-\lambda) &=&
%\kappa-\delta-T\left(2\frac{|s|}{|\psi|}
%+\frac{|\psi|s_{z}}{|s|}\right)\cos(\phi-\lambda)\\
\partial_{t} |\psi| & = & -2T|s_-|\sin(\varphi-\lambda), 
\\
\partial_{t}\varphi & = & \kappa-2T\frac{|s_-|}{|\psi|}\cos(\varphi-\lambda), 
\label{dotphi} 
\\
\partial_{t}\lambda & = & \delta
+T\frac{|\psi| s_{z}}{|s_-|}\cos(\varphi-\lambda). 
\label{dotlambda}
\end{eqnarray}
Up to now we have treated Eqs.\,\eqref{eq_cond} and 
\eqref{eq_dot} without any approximations.
In the following, we consider the limit  
$|s_-|^2/|\psi|^2 \ll 1$, 
which is natural given that we assume the condensates
to be sufficiently large for the Gross-Pitaevski\v\i\/ description to apply. 
Furthermore, we assume that $s_z$ oscillates around an average 
value $\bar s_z$, so that the average value of $|\psi|^2$ is
$C-\bar s_z$, accordingly. Similarly, we denote the average value
of $\partial_{t} \varphi$ by $\bar\omega$. 
Differentiating $|\psi|^2$ twice we obtain 
\begin{equation}
\frac{\partial_{t}^{2}|\psi|^{2} }{|\psi|^{2}}
%& = & 8T^{2}\frac{|s_-|^{2}}{|\psi|^{2}}+4T^{2}s_{z}
%-4T\frac{|s|}{|\psi|}(\kappa-\delta)\cos(\phi-\lambda)\\
 = 
 8T^{2}\frac{|s_-|^{2}}{|\psi|^{2}}+4T^{2}s_{z}
+2(\partial_{t}\varphi-\kappa)(\kappa-\delta).
\end{equation}
Neglecting the first term on the right hand side
and replacing $|\psi|^2$ and $\partial_t\varphi$ by their averaged values, 
we get
\begin{equation}
\partial_{t}^{2}|\psi|^{2}\approx4T^{2}(C-|\psi|^{2})(C-\bar{s}_{z})+2(\bar{\omega}-\kappa)(\kappa-\delta)|\psi|^{2}, 
\end{equation}
which results in the analytical solutions
 \begin{eqnarray}
|\psi|^{2} & = & 
C-\bar{s}_{z}-A_{0}\cos[2T\sqrt{C}(t-t_{0})],\nn 
s_{z} & = & \bar{s}_{z}+A_{0}\cos [2T\sqrt{C}(t-t_{0})],
\end{eqnarray}
and the self-consistency condition 
\begin{equation}
-2T^{2}\bar{s}_{z}=(\bar{\omega}-\kappa)(\kappa-\delta). 
\label{selfcons}
\end{equation}
The quantities $|\psi|^{2}$ and $s_z$
therefore oscillate  around their average values with the frequency 
$\omega_1 = 2T\sqrt{C}$. For the remaining determination of the 
constants $A_0, t_0,\bar s_z$ and
the average frequency $\bar \omega$, 
we use from now on the simplifying assumption of $\psi(0)>0$, i.e. $\varphi(0)=0$, 
and $s_{-}(0)$ being real. In that case, the amplitude is given by $A_0=s_z (0)-\bar s_z$,
whereas $t_0 =0$.
Moreover,  a good approximation for $\bar\omega$
is then given as the average of two extrema of $\partial_t\phi$ 
in Eq.\,\eqref{dotphi} according to 
$\bar{\omega}=\kappa- \textrm{sgn}[s_- (0)] T\left(\frac{|s_-(0)|}{|\psi(0)|}
+\frac{|s_-(\pi/2T\sqrt{C})|}{|\psi(\pi/2T\sqrt{C})|}\right)$. 
After a lengthy, but straightforward calculation we finally 
obtain
\begin{equation}
\bar{\omega}=\kappa- 2T^{2}\frac{(\kappa-\delta)s_{z}(0)
+2\omega_{1}s_-(0)}{(\kappa-\delta)^{2}+\omega_{1}^{2}},
\end{equation} 
which also directly determines $\bar{s}_z$  via Eq. (\ref{selfcons}). The final 
result can then be written as
\bea
\psi(t)
&=&\sqrt{C-\bar{s}_{z}-A_0 \cos(2T\sqrt{C}t)}\,e^{i\bar\omega t},
\nn 
\bar{\psi}(t)&=&\bar{\psi} (0)e^{-i\kappa t}\,,\quad \mbox{where} 
  \quad A_0= s_z(0)- \bar s_z. 
\ea 
Using that Re$[\psi\bar\psi^*]=|\psi_1|^2-|\psi_2|^2$, 
the particle imbalance oscillates according to 
\begin{multline}
%|\varPsi_{1}|^{2}-|\varPsi_{2}|^{2}   \\
n(t) =  \frac{\sqrt{C-\bar{s}_{z}-A_0\cos(\omega_1 t)}}{N_0}\, \textrm{Re}
\,[e^{i\Omega t} \bar{\psi}^*(0)].
\end{multline} 
It executes small oscillations of the amplitude at the frequency
$\omega_1 = 2T\sqrt{C}$, cf.\,\,Fig.\,\ref{small}, while the major
oscillations of the envelope are given by 
\bea \Omega & = & 2\kappa-
2T^{2}\frac{(\kappa-\delta)s_{z}(0)
+2\omega_{1}s_-(0)}{(\kappa-\delta)^{2}+\omega_{1}^{2}}. \label{Omega}
\ea 
When $T\rightarrow 0$, 
we have simply conventional Josephson oscillations at the frequency
$2\kappa$. Conversely, in the strong coupling limit, 
%i.e. for $\kappa$ negligibly small, 
and with one atom initially
in the dot, the envelope frequency $\Omega$ 
%of large amplitude oscillations 
depends in a simple way on
$\beta$: $\frac{\Omega}{T}=\frac{2\beta}{\beta^2+4C}$,
with a maximum at $\beta=2\sqrt{C}$.
% which is defined by initial
% conditions for the condensate wave-function and the occupation of the
% dot.
Comparing the frequency $\Omega$ with our numerical results shows excellent
agreement. For the set of parameters used in  
Fig.\,\ref{large}\,(a), for example, the analytically obtained periods
are  $284\,T^{-1}$ and  $52\,T^{-1}$ for  $\Gamma=0$ and $\Gamma=0.05$, respectively.
The first period corresponds exactly to the green curve. 
Comparing the latter with the red curve shows 
that the agreement remains  very good even for small 
interactions (finite $\alpha$), provided $\Gamma$ is also finite.
\bigskip 

%\vspace*{-1em} 
%\bc
%%%%%%%%%%%%%%%%%%%%%%%%%%%%%%%%%%%%%%%%%%%%%%%%%%%%%%%%%%%%%%%%%%%%%%%%%%%%%%
\begin{figure}[!hbt]
\bc
\includegraphics[angle=0,width=0.28\textwidth]{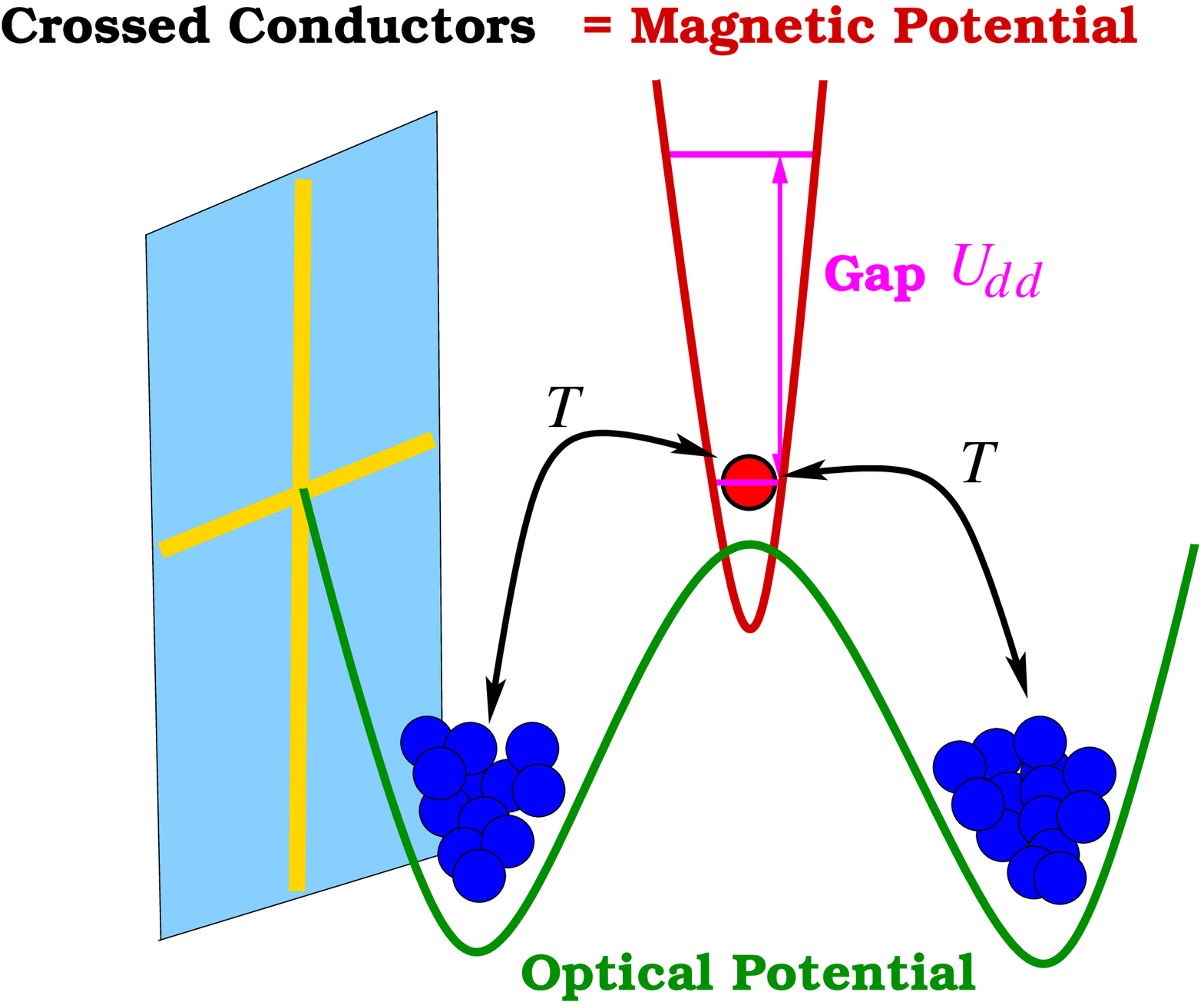}
%,height=0.2\textheight
%\vspace*{-1em}
\caption{\label{setup} (Color online) 
Possible experimental setup for an atomic quantum dot coupled to BECs
on a microchip (light blue).
A standing laser wave (green) creates deep potential wells 
perpendicular to the chip surface, equivalent 
to an optical potential for all hyperfine states of the atoms.
Two crossed conductors on the chip (yellow) 
create a tight magnetic potential 
for one hyperfine species (red) causing a large %(contact interaction) 
gap $U_{dd}$ 
for double occupation of the dot. An additional laser (not shown) couples
the impurity atom (red) and the condensates (dark blue), resulting in the 
transfer coupling $T$. 
}
\ec
\end{figure}
%%%%%%%%%%%%%%%%%%%%%%%%%%%%%%%%%%%%%%%%%%%%%%%%%%%%%%%%%%%%%%%%%%%%%%%%%%%%%%
%\ec

We finally discuss a possible 
experimental realization of our theoretical proposal. 
For that purpose, we suggest a combined 
microtrap--standing optical wave setup, 
in which a standing laser beam and crossed conductors on a 
microchip create the required trapping architecture 
\cite{Reichel,RMP}, see Fig.\,\ref{setup}.  
We note that all atoms %independent of their hyperfine state 
feel the optical potential used in this setup.  
%in the $m=-1$ state also feel the optical potential, 
However, the magnetic potential acting on one particular 
hyperfine state %, say $m=-1$, 
dominates the optical potential at the AQD location,  
provided that the ``contact interaction gap'' $U_{dd}$ (the energy 
barrier for double occupation of the dot) dominates the  
other relevant energy scales,
%: The Rabi frequency $\Omega_R$ 
%of the two-photon Raman transition coupling the $m=-1$ atoms into the dot
%and the chemical potential $\mu$ for the BECs inside the wells, i.e.
$U_{dd}\gg T , UN_0$. 

In conclusion, we have shown that
an atomic quantum dot can act as a coherent ``shuttle'' between
two essentially isolated BECs, transferring atoms from left to right and
vice versa such that Josephson-like oscillations are established. 
Using an analytical approximation in the noninteracting limit, we obtained
explicit expressions for the two frequencies characterizing the dot-induced 
oscillations. 
We have numerically established a phase diagram for non-self-trapped 
and self-trapped phases of the system, and analyzed experimental
feasibility.
% when self-interaction becomes relevant.
An extension of the present work 
is to consider arrays of atomic quantum dots
in optical lattices, where phenomena like self-trapping \cite{Anker} 
and the influence of the array on the Mott insulating state can be studied.  

\acknowledgments
We thank N. Schopohl, P. Treutlein, and I. Carusotto for helpful discussions. 
URF and AP acknowledge support by the Institut Henri Poincar\'e--Centre 
Emile Borel.

%\smallskip 
%$^*$\,{\footnotesize \sf uwe.fischer@uni-tuebingen.de}     
%$^\S$\,{\footnotesize\sf anna@tfp.physik.uni-karlsruhe.de} 


\begin{thebibliography} {99}

\bibitem{Josephson} B. D. Josephson,
%: {\em Possible new effects in   superconductive tunneling}, 
Phys. Lett. {\bf 1}, 251 (1962).

\bibitem{Pereverzev} S. V. Pereverzev, A. Loshak, S. Backhaus, J. C. Davis,
and  R. E. Packard,
%: {\em Quantum oscillations between two weakly coupled   reservoirs of superfluid $^3$He}, 
Nature {\bf 388}, 448 (1997). 

\bibitem{MilburnJosephson} G.\,J. Milburn, J. Corney, 
E.\,M. Wright, and D.\,F. Walls,
%: {\em  Quantum dynamics of an atomic Bose-Einstein condensate in a double-well potential}, 
Phys. Rev. A {\bf 55}, 
4318 (1997).

\bibitem{Smerzi} A. Smerzi, S. Fantoni, S. Giovanazzi, and S. R. Shenoy, 
%: {\em Quantum Coherent Atomic Tunneling between Two Trapped Bose-Einstein Condensates}, 
Phys. Rev. Lett. {\bf 79}, 4950 (1997).

\bibitem{Zapata} I. Zapata, F. Sols, and A.\,J. Leggett,
%: {\em Josephson effect between trapped Bose-Einstein condensates}, 
Phys. Rev. A {\bf 57},  R28 (1998). 

\bibitem{Raghavan} S. Raghavan, A. Smerzi, 
S. Fantoni, and S.\,R. Shenoy, Phys. Rev. A {\bf 59}, 620 (1999). 

\bibitem{Albiez} M. Albiez, R. Gati, J. F\"olling, 
S. Hunsmann, M. Cristiani, and M.\,K. Oberthaler, 
%: {\em Direct Observation of  Tunneling and Nonlinear Self-Trapping in a single Bosonic Weak Link}, 
Phys. Rev. Lett. {\bf 95}, 010402 (2005). 

\bibitem{Levy} S. Levy, E. Lahoud, I. Shomroni, and J. Steinhauer, 
%"The a.c. and d.c. Josephson effects in a Bose-Einstein condensate", 
Nature {\bf 449}, 579 (2007).

\bibitem{Recati} A. Recati, P.\,O. Fedichev, W. Zwerger, J. von Delft, 
and P. Zoller,
%: {\em Atomic Quantum Dots Coupled to a Reservoir of a Superfluid 
%Bose-Einstein Condensate}, 
Phys. Rev. Lett. {\bf 94}, 040404 (2005). 

\bibitem{Bausmerth} I. Bausmerth, U.\,R. Fischer, and A. Posazhennikova, 
%: {\em Quantum top inside a Bose-Einstein-condensate Josephson junction}, 
Phys. Rev. A {\bf 75}, 053605 (2007). 

%\bibitem{Shnirman}  J.-X. Zhu, Z. Nussinov, A. Shnirman, and A.\,V. Balatsky, 
% Phys. Rev. Lett. {\bf 92}, 107001 (2004).  

%\bibitem{zapataII}  I. Zapata, F. Sols, and A.\,J. Leggett,
%: {\em Phase dynamics after connection of two separate Bose-Einstein condensates}, 
%Phys. Rev. A {\bf 67}, 021603(R) (2003).


%\bibitem{leggettBEC} A.\,J. Leggett, 
%: {\em Bose-Einstein condensation in the alkali gases: Some fundamental concepts}, 
%Rev. Mod. Phys. {\bf 73}, 307 (2001).
% [Erratum: {\it ibid.} {\bf 75}, 1083 (2003)].


\bibitem{Roati} Feshbach resonances have recently 
been used to approach the noninteracting BEC limit 
by G. Roati {\it et al.}, Phys. Rev. Lett. {\bf 99}, 
010403 (2007). 

\bibitem{Reichel} J. Reichel, 
%: {\em Microchip traps and Bose-Einstein condensation}, 
Appl. Phys. B {\bf 74}, 469 (2002).  

\bibitem{RMP} J. Fort\'agh and C. Zimmermann,
%: {\em Magnetic microtraps for ultracold atoms}, 
Rev. Mod. Phys. {\bf 79}, 235 (2007). 

\bibitem{Anker} Th. Anker {\it et al.}, 
%M. Albiez, R. Gati, S. Hunsmann, B. Eiermann, A. Trombettoni, and M.\,K. Oberthaler, 
Phys. Rev. Lett. {\bf 94}, 020403 (2005).  
  
\end{thebibliography}
\end{document}